# Networks of superconducting nano-puddles in 1/8 doped $YBa_2Cu_3O_{6.5+y}$ controlled by thermal manipulation

Alessandro Ricci[1,2], Nicola Poccia[2,3], Gaetano Campi[4], Francesco Coneri[3], Luisa Barba[5], Gianmichele Arrighetti[5], Maurizio Polentarutti[5], Manfred Burghammer[6,7], Michael Sprung[1], Martin v. Zimmermann[1], Antonio Bianconi[2,4]

[1]*Deutsches Elektronen-Synchrotron DESY, Notkestraße 85, D-22607 Hamburg, Germany*
[2]*Rome International Center for Materials Science Superstripes RICMASS, via dei Sabelli 119A, 00185 Roma, Italy*
[3]*MESA+ Institute for Nanotechnology, University of Twente, P. O. Box 217, 7500AE Enschede, Netherlands*
[4]*Institute of Crystallography, CNR, via Salaria Km 29.300, Monterotondo Roma, I-00015, Italy*
[5]*Elettra Sincrotrone Trieste. Strada Statale 14 - km 163, 5, AREA Science Park, 34149 Basovizza, Trieste, Italy*
[6]*European Synchrotron Radiation Facility, B. P. 220, F-38043 Grenoble Cedex, France*
[7]*Department of Analytical Chemistry, Ghent University, Krijgslaan 281, S12 B-9000 Ghent, Belgium*

While it is known that the nature and the arrangement of defects in complex oxides have an impact on the material functionalities little is known on control of superconductivity by oxygen interstitial organization in cuprates. Here we report direct compelling evidence for the control of Tc, by manipulation of the superconducting granular networks of nanoscale puddles, made of ordered oxygen stripes, in a single crystal of $YBa_2Cu_3O_{6.5+y}$ with average formal hole doping p close to 1/8. Upon thermal treatments we were able to switch from a first network of oxygen defects striped puddles with OVIII modulation ($q_{OVIII}(a^*)$=(h+3/8,k,0) and $q_{OVIII}(a^*)$=(h+5/8,k,0)), to second network characterized by OXVI modulation ($q_{OXVI}(a^*)$=(h+7/16,k,0) and $q_{OXVI}(a^*)$=(h+9/16,k,0)), and finally to a third network with puddles of OV periodicity ($q_{OV}(a^*)$=(4/10,1,0) and $q_{OV}(a^*)$=(6/10,1,0)). We map the microscopic spatial evolution of the out of plane OVIII, OXVI and OV puddles nano-size distribution via scanning micro-diffraction measurements. In particular, we calculated the number of oxygen chains (n) and the charge density (holes concentration p) inside each puddle, analyzing areas of 160x80 µm$^2$, and recording 12800 diffraction patterns to reconstruct each spatial map. The high spatial inhomogeneity shown by all the reconstructed spatial maps reflects the intrinsic granular structure that characterizes cuprates and iron-chalcogenides, disclosing the presence of several complex networks of coexisting superconducting domains with different lattice modulations, charge density and different gaps like in the proposed multi-gaps scenario called superstripes.

## 1. Introduction

An essential step towards the understanding of modern materials and their implementation in novel nano-electronic devices is the control and manipulation of their microscopic behavior [1-3]. Recently, the interrelationship between spin, charge, and lattice orders in high temperature superconductors (HTS) is at the center of a very animated discussion [4-15]. Novel results obtained in $YBa_2Cu_3O_{6+y}$ (YBCO), provide compelling evidence for the charge density waves (CDW), and static magnetic stripes are intertwined and aggregated in nanoscale puddles [16-22]. These domains are spatially separated by superconducting regions composed by ordered lattice stripes [23-27] forming an intrinsically complex lattice of striped puddles called "superstripes" scenario [13]. In this scenario the local lattice modulations determine multiple subbands crossing the Fermi level and therefore multi-gap superconductivity below the critical temperature [28-29]. This theoretical proposal has been recently supported by the prediction of the anomalous isotope coefficient at 1/8 doping [30]. Since defects nature and distribution becomes the driving force it is of high relevance their control by thermal annealing. Therefore a primary task for both fundamental physics and novel nano-electronics is the careful visualization of the effects of a thermal treatment on the system, via imaging the quasi 2-dimensional puddles of oxygen chains in YBCO. Unfortunately, due to the lack of proper local bulk-sensitive probes, the microscopic scenario is still not clear and the real-space and real-time observation of thermally induced rearrangements of superconducting micro-regions in HTS is a difficult experimental task.

The development of a technique for imaging the nanoscale intrinsic inhomogeneity of oxygen chains organization is the first step to open new opportunities for their manipulation. Here we explore the nanoscale granular patterns arising in $YBa_2Cu_3O_{6.67}$ single crystal with doping close to 1/8 holes content per Cu site in the $Y(CuO_2)_2$ bilayer. The $YBa_2Cu_3O_{6.67}$ crystal is an ideal system for the investigation of inhomogeneity due to the short range ordering of oxygen ions [31] at 1/8 doping. Indeed, it exhibits an incommensurate superlattice of (OVIII) chains indicated by the lattice superstructure modulation at $q_{OVIII}(a^*)=(h+3/8, k, 0)$ and $q_{OVIII}(a^*)=(h+5/8, k, 0)$. Although the oxygen tendency to form O-Cu-O fragments in the basal plane has been widely investigated [32,33], to date there is little information on the spatial distribution of these fragments, and on their inclination towards aggregation in domains.

In the first part of our paper we report the temperature evolution in $YBa_2Cu_3O_{6.67}$ using standard synchrotron X-ray diffraction of the OVIII puddles of oxygen chains. Using the heat treatment we show the formation of a second network of puddles characterized by OXVI superstructure modulation at $q_{OXVI}(a^*)= (h+7/16,k,0)$ and $q_{OXVI}(a^*)=(h+9/16,k,0)$, and finally a third network of OV puddles characterized by the superstructure wavevectors $q_{OV}(a^*)=(4/10,1,0)$ and $q_{OV}(a^*)=(6/10,1,0)$. The measurements provide us information about the average order of oxygen chains in the sample. In the following, we investigate the dynamics and spatial distribution of OVIII, OXVI and OV domains upon thermal cycling via micro X-ray diffraction (μXRD). By scanning micro areas, this technique provides mixed information of the reciprocal and real-space of the bulk structure inhomogeneities, and it has never been applied on a $YBa_2Cu_3O_{6.67}$ single crystal so far. Finally we show how through thermal treatment it is possible to control the puddles size distribution, the number of oxygen chains and their charge density. Indeed using the novel experimental method μXRD has been possible to directly visualize how the thermal treatment affects the intrinsic nanoscale phase heterogeneity in this cuprate superconductor. We relate these changes with the onset variation of the superconducting temperature that we readily can control over a range of 2K.

2. **Materials and methods**

Starting reagents with ultra-high purity (CuO and $Y_2O_3$ 99.999%, $BaCO_3$ 99.997%) have been employed to grow extremely good single crystals of $YBa_2Cu_3O_{6.67}$ into Barium Zirconate crucibles by the self-flux technique [34]. Inductively coupled plasma mass spectroscopy indicates a purity of the crystals higher than 99.99 atomic percentage. The oxygen content of the crystal was set to 6.67 by annealing in flowing oxygen at 914°C, followed by quenching to room temperature under flowing nitrogen gas. The macroscopic oxygen content inhomogeneity was removed by annealing the crystal at 570 °C in a sealed quartz capsule, followed by quenching in an ice-water bath. The crystal was then kept at room temperature to let the short range oxygen order establish. Average structure characterization show a unit cell described by a P4/m spatial symmetry and lattice dimensions: a=3.807(11) Å, b=3.864(12) Å, c=11.52(2) Å, with a unit cell volume of 169.5(8) Å$^3$.

Temperature dependent X-ray diffraction measurements were performed on the XRD1 beamline at the ELETTRA synchrotron storage ring, Trieste. The beamline is placed on a multipole wiggler insertion device operating under the current ELETTRA conditions of 2 GeV ring energy and 400 mA injection current (see Fig. S1). The samples were oriented on a kappa diffractometer equipped with a motorized goniometric X-Y stage head and a Mar-Research 165 mm CCD camera as detector. The data were collected in transmission mode with a photon energy of 20 KeV (λ= 0.61992 Å), selected from the source by a double-crystal Si(111) monochromator and using a beam of $200\times200\mu m^2$. The 2D CCD detector (MAR-Research) was placed at a distance of 70 mm from the sample. Data from a $LaB_6$ standard were collected for calibration. Measurements were conducted between 300 and 400 K with a temperature step of 2 K for both the heating (300 to 400 K) and the cooling (400 to 300 K) cycles. Temperature was varied and controlled by means of a cryo-cooler 700 series Oxford Cryosystems that allows working in a range of 90K-400K guaranteeing accuracy better than ±1 K. All the images measured by the single crystal x-ray diffraction were processed using FIT2D program jointly to a MATLAB® based software-package developed in-house. Secondly, the microscopic behavior of the sample under the thermal treatments has been investigated using scanning μXRD in reflection geometry (on *a-c* plane).

The ID13 beam line of the European Synchrotron Radiation Facility (ESRF) is specialized in the delivery of micro-focused x-ray beams for x-ray diffraction experiments. The photon source, in the range 12-13 KeV, is a 18 mm period in vacuum undulator at the ESRF 6.03 GeV storage ring operated in multi-bunch mode with a current of 200 mA. The optics of the micro-focus beamline include compound refractive lenses, Kirkpatrick Baez (KB) mirrors, crossed Fresnel zone plates or waveguides. The ellipsoidal mirror is the main focusing element, demagnifying the source by a factor of 10 (about 40 microns in diameter). The focused beam is defined by a pinhole of 5 micron diameter. The beam is focused by a tapered glass capillary to 1 micron in diameter. The beam-line uses two monochromators positioned in series; the first is a liquid $N_2$ cooled Si-111 double crystal or Si-111 (bounce); the second, is a channel cut monochromator employing a single liquid nitrogen cooled Si crystal. The detector of x-ray diffraction images is a high resolution CCD camera (Mar CCD) with point spread function 0.1 mm, 130 mm entrance window, 16 bit readout placed at a distance of about 90 mm from the sample. To scan the sample area it has been moved using two piezoelectric stages in x-y direction and data have been collected in the θ-2θ reflection geometry (see Fig. S2). The huge amount of diffraction patterns collected by μXRD (more than 12800 each map) have been processed using a MATLAB® based software-package developed in-house.

The magnetic behavior vs. temperature of our YBCO sample at 1/8 has been characterized by means of the Vibrating Sample Magnetometer (VSM) option in a Physical Properties Measurement System (PPMS 6000) from Quantum Design. Here a linear motor vibrates the sample with a frequency of 40 Hz and amplitude of 2 mm at the center of a pick-up coil, and the induced voltage is measured synchronically with the oscillation. A magnetization measurement has been chosen to consist of the averaged value over 40 data points that is over a period of one second of oscillation. These parameters guarantee a good signal to noise ratio.

### 3. Results and Discussion

The ordering process of the oxygen ions in chains domains has been studied using transmission XRD. The diffraction pattern due to the superlattice reflections was recorded at room temperature and shown in Figure 1a.

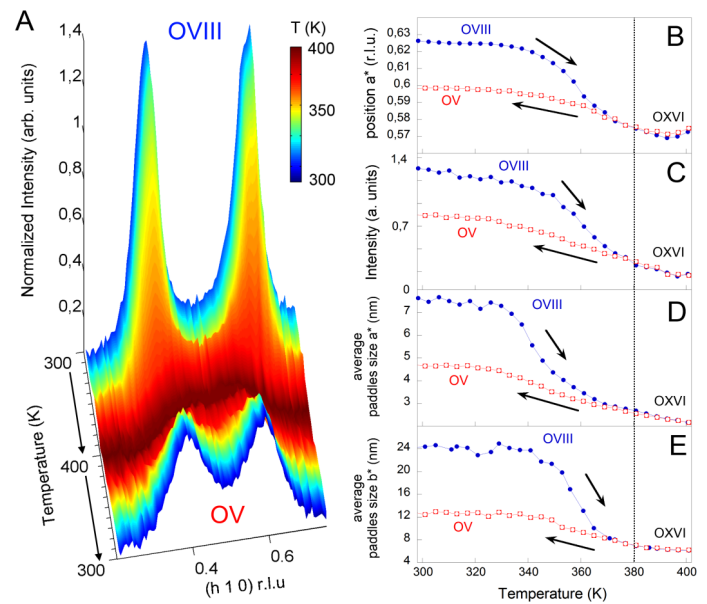

**Figure 1.** (colored online): **(a)**. Three dimensional color plot of the thermal evolution of the OVIII oxygen stripes superlattices profile at wavevectors $q_{OVIII}(a^*)=(3/8, 1, 0)$ and $q_{OVIII}(a^*)=(5/8, 1, 0)$ into the OV superstructure at wavevector $q_{OV}(a^*)=(4/10, 1, 0)$ and $q_{OV}(a^*)=(6/10, 1, 0)$. The temperature ranges from 300 K to 400 K which respectively corresponds to dark blue and red colors. The reflections of the superstructures in the $YBa_2Cu_3O_{6+y}$ crystal are measured using a large ($200 \times 200 \mathrm{\mu m}^2$) beam at the ELETTRA storage ring. **(b)** The thermal cycle of the superlattice position along $a^*$ crystallographic direction, **(c)** the intensity, **(d)** the average puddles size along $a^*$ **(e)** and along $b^*$. The arrows show the temporal direction of the experiment. The blue filled circles show the evolution from room temperature to 400 K. The red empty squares show the evolution from 400 K to room temperature. The black dotted-line indicates the temperature of 380 K, ones crossed it the OVIII order can be recovered only waiting few days at room temperature.

It is possible to distinguish superlattice peaks due to the OVIII (6/16,1,0) and (10/16,1,0) reflections in the a*b* plane at 300 K in agreement with previous data [22] which confirms the high quality of our single crystal. We monitored the X-ray diffraction patterns during a thermal cycle from 300 K to 400 K, and then back to 300 K, using a slow rate of 0.5 K/min. The continuous evolution of the superlattice profiles from OVIII to OV through OXVI modulation is shown in **Figure 1b**. The OVIII (6/16,1,0) and (10/16, 1, 0) reflections in the a*b* plane at 300 K evolve into the OXVI (7/16,1,0) and (9/16,1,0) modulation upon heating up to 400K.

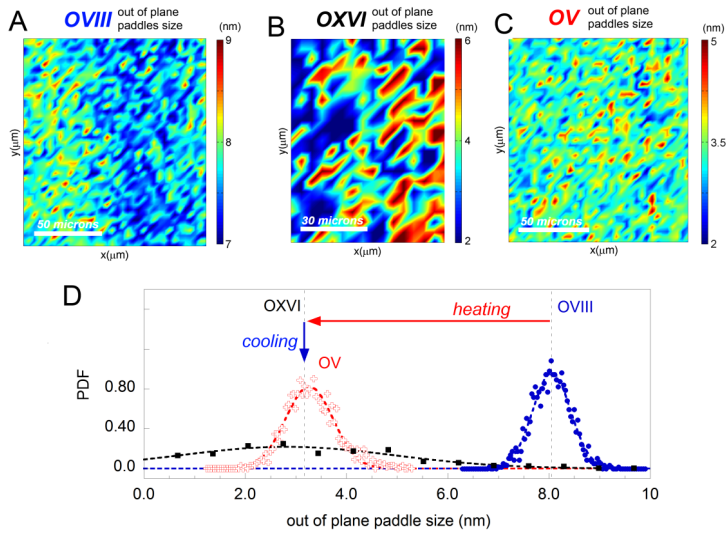

**Figure 2.** (colored on-line): The out of plane puddles sizes (puddles sizes along c*) before the heating cycle **(a)** at 390K **(b)** and at the end of the cooling cycle **(c)** in the OVIII, OXVI and OV phase, respectively. **(d)** The probability density function of the OVIII (blue filled circles), OXVI (black filled squares) and OV (red empty crosses) out of plane puddles sizes. The out of plane puddles sizes decrease after the thermal heating cycle from OVIII to OXVI where their distribution looks very broad. After the cooling cycle the average out of plane puddles sizes remain the same in the OV phase. On the other hand, the shape of the distribution becomes much sharper, indicating a certain recovered order. Dotted lines are guides for the eyes.

During the subsequent cool down to 300 K the original OVIII modulation is not restored and a new OXVI phase, identified by the superlattice peaks (7/16, 1, 0) and (9/16,1,0), sets in. In **Figure 1b**, **1c**, **1d** and **1e** we report respectively the temperature dependence of the superlattice peaks position, the intensity, and the average domain size along a* and b*. The average domain size along the in-plane a* and b* directions have been calculated fitting the superlattice reflections by a two-Lorentians model. In addition, above 380 K the system crosses a OXVI phase with wavevector $q_{mix}=(h\pm 9/16,k,0)$. The average size of the OVIII domain along a* is of about 7.5(5) nm, but approaching the OV, it decreases down to 4.5(5) nm. The average size is larger along b*, showing OVIII and OV domains of about 23.5(5) nm and 11.5(5) nm, respectively. This phenomenology shows that the in plane average domain size of oxygen chains can be controlled and manipulated by tuning the temperature in a quasi-irreversible manner. Leaving the sample under vacuum at room temperature, more than one month is needed for the OV phase to spontaneously drive itself back to a new reconstructed OVIII phase.

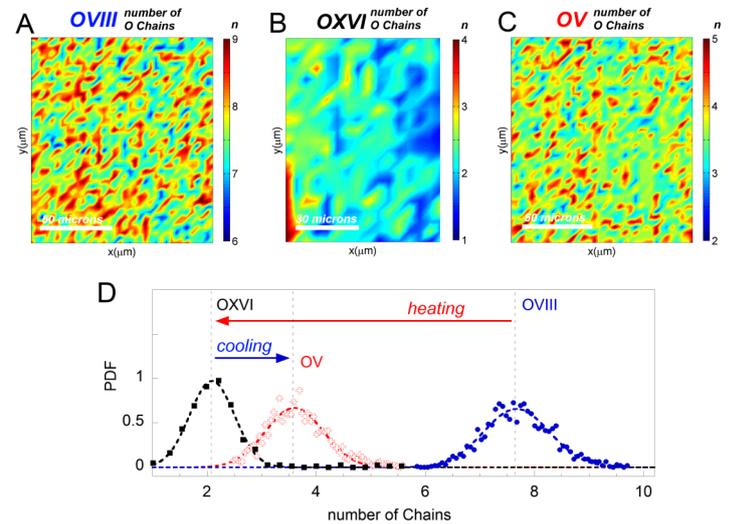

**Figure 3.** (colored on-line): **(a)**, **(b)** and **(c)** The spatial map of number of the oxygens chains inside the OVIII, OXVI and OV puddles, have been reported. **(d)** Distribution of the number of oxygens chains into the OVIII (blue filled circles), OXVI (black filled squares) and OV (red empty crosses) puddles. The number of oxygens chains decreased after the thermal heating cycle from OVIII to OXVI and increased again after the cooling in the OV puddles but remain still lower than in the OVIII phase. Dotted lines are guides for the eyes.

This aspect shows relevant analogies with the oxygens and local lattice distortions ordering in the $La_2CuO_{4+y}$, where the Q2-iO phase (due to interstitial oxygens ordering) and the Q3-LLT (due to local lattice distortions ordering) can be changed by thermal treatments and restored waiting a long time or by the use of X-ray continuous illumination [35-40].
In order to monitor the microscopic evolution of OVIII domains into the OXVI and OV during the thermal manipulation, we used μXRD like already successfully done to investigate phase separation appearing in other cuprates [35-40] and iron-based superconductors [41-44]. We performed our measurements using a beam of about 1 μm of diameter. The sample has been scanned over an area of 160x80 μm² collecting 12800 micro-diffraction patterns showing superlattices at $q_{OVIII}(a^*)=(3/8,0,4)$ and $q_{OVIII}(a^*)=(5/8,0,4)$ due to OVIII puddles of ordered oxygen chains. The superlattice peaks profiles have been fitted using a 2-Lorentians model and the satellite position $h_{XY}$ along a* and the FWHM$_{XY}$ along a* and c*, have been extracted for every micro-diffraction pattern at X-Y position.

From the FWHM$_{XY}$ along c* we calculated the out of plane domain size for every spot of the scanned map. **Figure 2a**, **2b** and **2c** show the real space map of the out of plane domain size (along c*), before the heating (OVIII), at 390K (OXVI) and after the cooling cycle (OV), respectively. These maps show the presence of clear intrinsic nanoscale heterogeneity in this cuprate superconductor like has been observed in iron-chalcogenides [41,43].

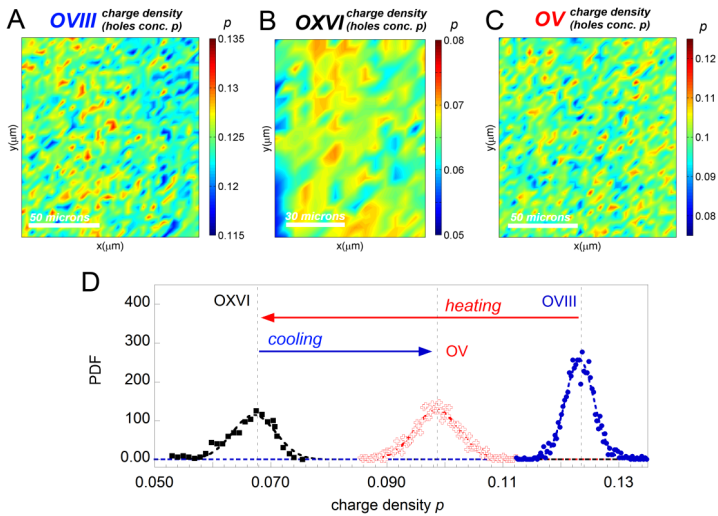

**Figure 4.** (colored on-line): **(a)**, **(b)** and **(c)** The spatial map of the charge density (holes concentration p) of the OVIII, OXVI and OV puddles, have been reported. **(d)** Distribution of the charge density into the OVIII (blue filled circles), OXVI (black filled squares) and OV (red empty crosses) puddles. During the thermal annealing process the charge density changed and its distribution in the OXVI and OV puddles get broader.

The OVIII probability density functions for the domain sizes distribution along c* is relatively sharp, and varies from 7 to 9 nm. Upon heating, the OXVI phase shows instead a very broad distribution, that spans the 1 to 6 nm range. After the cooling cycle, the size distribution in the OV phase sharpens again, around the decreased average value of 3 nm, therefore indicating a certain recovered order. The overall behavior indicates a sensible response of the granular network to thermal treatment.

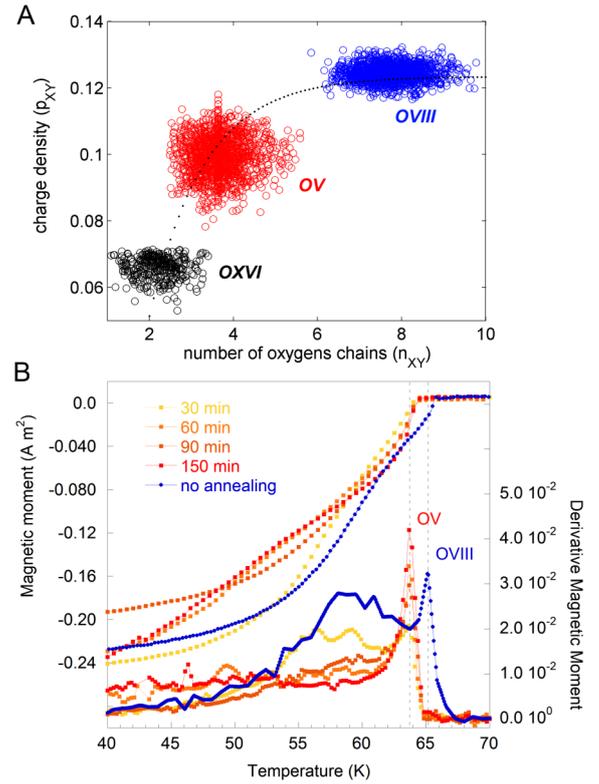

**Figure 5.** (colored on-line): **(a)** Charge density (holes concentration p) as a function of the number of oxygens chains, inside the OVIII, OXVI and OV puddles. The dotted curve is a fit using a model: $p = 1 - e^{-[(n-n_0)/n_0]/\hat{\imath}_n}$. Where $n_0$ and $\hat{\imath}_n$ are the minimum and the maximum number of chains present in the average puddle. **(b)** left scale: Zero Field Cooling (ZFC) diamagnetic response of YBCO upon thermal cycles in an external applied field H=20 Oe. Blue filled circles: signal before any thermal annealing. Yellow-red squares: the magnetic moment upon sample annealing at 380K with increasing dwell time of 30, 60, 90 and 150 minutes. Right scale: numerical derivative of the magnetic moment. Upon thermal cycling the onset of superconducting shielding decreases of about 2K.

Starting from the FWHM$_{XY}$ and the $h_{XY}$ along a*, measured at each X-Y spatial position, we reconstructed the spatial maps of the number of oxygen chains (n) inside the puddles. Using the expression: $n_{XY} = (1 - h_{XY})/FWHM_{XY}$, we calculate this quantity for the OVIII **(Figure 3a)**, OXVI **(Figure 3b)** and for the annealed OV, puddles **(Figure 3c)**.

The number of oxygen chains decreases after the thermal heating cycle from OVIII to OXVI and increases again after the cooling in the OV puddles (but remains still lower than in the OVIII phase).

**Figure 4a**, **4b** and **4c** show the spatial map of the charge density (or holes concentration p) inside the puddles of oxygen chains, onto the same area described in **Figure 3**, before the heating (OVIII phase), at 390K (OXVI phase) and after cool down (OV phase). The charge density has been calculated considering the difference of the superlattices position along a* ($h_{XY}$) with respect to the ortho-II (OII) phase following the relationship: $p_{XY} = h_{XY} - 0.5$. Where the OII modulation corresponds with a periodicity of a filled CuO chain intercalated by one empty chain. During the thermal annealing process the charge density changed on the microscopic scale and its distribution in the OXVI and OV puddles gets broader, demonstrating the strong granularity of the system. **Figure 5a** shows the spot to spot charge density ($p_{XY}$) as a function of the number of oxygen chains ($n_{XY}$), inside the OVIII, OXVI and OV puddles. The behavior has been fitted using an exponential model: $p_{XY} = 1 - e^{-[(n_{XY} - n_0)/n_0]/\hat{i}_n}$.

Here $n_0$ and $\hat{i}_n$ are the minimum and the maximum number of chains present in the average puddle. In order to understand how the microscopic reduction of the effective hole-doping affects the superconducting properties, we studied the magnetic response of our sample across the superconducting transition before and after the thermal annealing, i.e. in the OVIII and in the OV phase, by means of a Vibrating Sample Magnetometer (VSM) option in a Physical Properties Measurement System (PPMS 6000) from Quantum Design (see supplementary info). The results are shown in Figure 5b. A first measurement is performed before any annealing procedure is carried on (OVIII phase), and shows the onset temperature of the diamagnetic screening ($T_c$) to be about 66 K. The subsequent measurements are performed upon thermal annealing at 380 K (OV phase), with increasing dwell time of 30, 60, 90 and 150 minutes. In all these cases $T_c$ decreases of about 2 K. This effect is irrespective of the annealing time, and we associate such a reduction with the lower effective charge density we point out in fig 4 and 5a.

## 4. Conclusion

In conclusion, we have investigated how thermal treatments allow to microscopically manipulate and control the functional properties of $YBa_2Cu_3O_{6.67}$ (p≈1/8). We used an X-ray diffraction approach, combining standard synchrotron XRD measurements ($200 \times 200 \mu m^2$ beam size) with scanning μXRD ($1 \times 1 \mu m^2$) and VSM. Using thermal annealing we induced a continuous phase transition that led to a different final arrangement of Cu-O chains into the sample. In particular we monitored a transition from the OVIII to the OV modulation for the oxygen chains domains, by cycling between 300K and 400K. The microscopic dynamics of the domains have been investigated by scanning μXRD. We mapped with micrometric resolution the out of plane domain size, the number of oxygen chains and the charge density inside each domain, covering a total area of 160x80μm². We recorded 12800 diffraction patterns for each spatial map, showing a high nanoscale inhomogeneity and the presence of a complex network-like organization of competing superconducting puddles that are characterized by different number of oxygen chains and charge density. A reduction in the out of plane domains size, in the number of oxygen chains and in the microscopic distribution of charge density have been observed in the OV phase. These reductions have been connected to a decrease of number of holes in the active layer. As a consequence, magnetization measurements show that the modification of the network structure of superconducting grains is responsible of a drop of $T_c$ of about 2 K. This can open the way to a possible $T_c$ tuning by microscopic thermal-manipulation of oxygens chains distribution in HTS. This work shows the presence of a microscale phase separation in $YBa_2Cu_3O_{6.67}$ with a hole doping close to 1/8 where the lattice misfit strain [44] in these heterostructures at atomic limit and the proximity of the Fermi level to a 2.5 Lifshitz transition near a band edge of the subbands [45,46,28-30] induce the observed nanoscale phase separation predicted by the multiband Hubbard model [47]. Finally in this experiment we observe a superstripes [13] lattice scenario in $YBa_2Cu_3O_{6.67}$ made of different striped nanoscale puddles of locally ordered interstitials with well defined hole doping density. The unique information that our experiment provides on the density distribution of the nanoscale striped puddles shows complex networks of superconducting units that supports they the statistical physics theories of percolative superconductivity in complex networks as an essential feature for understanding the emerging high temperature superconductivity [48-52]. In fact the reconstructed spatial maps shown here provide compelling evidence for the generic granular structure that characterizes cuprates and iron-chalcogenides. We disclose practical multiple realizations of complex networks of dopants self-organization at the nanoscale with striped puddles characterized by different modulations, local charge density and superconducting condensates which share the common features for the emerging of high Tc superconductivity.

## Acknowledgments

We thank Ruixing Liang, D. A. Bonn, and Walter N. Hardy of the Department of Physics of the University of British Columbia for providing us with the crystals and for helpful discussions. We thank the ID13 beamline staff of ESRF and the XRD1 beamline staff of ELETTRA, especially G. Bais. N.P. acknowledges the Marie Curie Intra European Fellowship for financial support.